\documentclass[aps,pra,showpacs,twocolumn,amsmath]{revtex4}
\usepackage{graphicx}

\newcommand{\beq}{\begin{equation}}
\newcommand{\eeq}{\end{equation}}
\newcommand{\beqa}{\begin{eqnarray}}
\newcommand{\eeqa}{\end{eqnarray}}
\newcommand{\bsub}{\begin{subequations}}
\newcommand{\esub}{\end{subequations}}

\newcommand{\rem}[1]{}
\newcommand{\refe}[1]{Eq.~(\ref{#1})}

\newcommand{\LLf}{{\cal L}_{\rm f}}
\newcommand{\LLs}{{\cal L}_{\rm s}}

\newcommand{\notes}[1]{}

\begin{document}
\title{Bistability of a slow mechanical oscillator coupled to a laser-driven two-level system}
\author{Fabio Pistolesi}
\affiliation{
Univ. Bordeaux, CNRS, LOMA, UMR 5798, F-33405 Talence, France
}
\begin{abstract}
It has been recently proposed that single molecule spectroscopy could
be employed to detect the motion of nano-mechanical resonators.
Estimates of the coupling constant ($g$) between the molecular 
two-level system and the oscillator indicate that it can 
reach values much larger than
the mechanical resonating pulsation ($\omega_m$) 
and the two-level system linewidth ($\Gamma$).
Other experimental realization of the same system are also 
approching this strong coupling regim.
In this paper we investigate the behavior of the system in the 
limit for slow mechanical oscillator $\omega_m \ll \Gamma$.
We find that, for sufficiently large coupling, the system undergoes a 
bistability reminiscent of that observed in optical
cavities coupled to mechanical resonators.
\end{abstract}

\rem{ PACS

81.07.Oj 	Nanoelectromechanical systems (NEMS)

42.50.Hz 	Strong-field excitation of optical transitions in quantum systems; multiphoton processes; dynamic Stark shift (for multiphoton ionization and excitation of atoms and molecules, see 32.80.Rm, and 33.80.Rv, respectively)

}

\date{\today}


 \maketitle

\section{Introduction}

The rapid developpement of nano-electromechanics in the last decade 
has seen the proposal and the experimental realisation of several systems
where in order to detect the displacement of a mechanical resonator it is coupled 
to a two-level system (TLS).
This includes superconducting Qbits
\cite{oconnell_quantum_2010},
nitrogen vacancy centers in diamonds 
\cite{arcizet_single_2011,rohr_synchronizing_2014,lee_deterministic_2014,
ovartchaiyapong_dynamic_2014,
reserbat-plantey_electromechanical_2016,pigeau_observation_2015,
lepinay_universal_2017,de_assis_strain-gradient_2017},
semiconductor TLS \cite{yeo_strain-mediated_2013, montinaro_quantum_2014,auffeves_optical_2014}, 
spins \cite{rabl_strong_2009,rabl_quantum_2010,rabl_cooling_2010,treutlein_hybrid_2014,
kolkowitz_coherent_2012}
or single molecules \cite{puller_single_2013}.
One of the interest in coupling a mechanical resonator to a TLS is that one can reach large 
couplings constants \cite{rabl_strong_2009,arcizet_single_2011,puller_single_2013,elouard_quantum_2015}.
According for instance to the estimates of Ref.~\cite{puller_single_2013}, 
the coupling constant can become larger than the mechanical pulsation
$\omega_m$ or the TLS linewidth $\Gamma$. 
Increasing the coupling not only improves the detection sensitivity of the mechanical 
displacement, but allows to reach new regimes, where the dynamics of the 
TLS and of the mechanical oscillator have to be considered on the same footing. 
This is well known for the case of mechanical resonators coupled to optical cavities \cite{marquardt_optomechanics_2009,aspelmeyer_cavity_2014,teufel_sideband_2011}, 
for which it is possible to reach the strong coupling limit since the effective 
opto-mechanical coupling constant is proportional to the square root of the 
number of photons present in the cavity.
A striking effect is the onset of a static bistability, that was observed 
long time ago \cite{dorsel_optical_1983,
gozzini_light-pressure_1985,aspelmeyer_cavity_2014}.
Reaching such a strong coupling limit with the bare coupling between the oscillator and the quantum system, in our case the TLS, is difficult, but currently at reach of the present technology.

In this paper we consider the case of a slow oscillator $\omega_m \ll \Gamma$
coupled to a laser-driven TLS and, by exploiting an adiabatic expansion, we obtain
a description of the TLS-oscillator system in the strong coupling limit.
We find that a static bistability is also expected, with a behaviour similar to that observed 
for optical cavities. 
In this case it is induced by the coupling to a single quantum degree of freedom 
of the TLS instead of the macroscopic condensate of photons.
Note alo that the presence of the TLS renders the problem intrinsically non-linear.
We calculate the luminescence that is the typical observed quantity in single-molecule spectroscopy experiments \cite{tamarat_ten_2000}.
We find that the interplay of the cooling-heating effect with the bistability gives rise to 
anomalous line-shapes.
This bistability resembles the one observed in optical cavities, but with the 
notable difference that the quantum nature of the TLS has to be taken into account. 

The paper is organized as follows. 
In Section \ref{sec:system} we present the model.
In Section \ref{sec:BornMarkov} the Born-Markov equations are derived.
In Section \ref{sec:adiabatic} we exploit the separation of time scales to 
simplify the Bork-Markov equations and obtain a description of the slow degree of freedom.
In Section \ref{sec:EffTemp} we discuss the effective temperature 
induced by the coupling to the TLS.
In Section \ref{sec:mechanicalbistability} the condition for the appearance of the mechanical bistability is discussed.
In Section \ref{sec:stochastic} the effect of the stochastic fluctuations is considered by solving numerically the Fokker-Planck equation. 
Section   \ref{sec:conclusions} gives our conclusions.

\section{System}
\label{sec:system}

\newcommand{\LL}{{\cal L}}

\begin{figure}[tbp]
  \includegraphics[width=8cm]{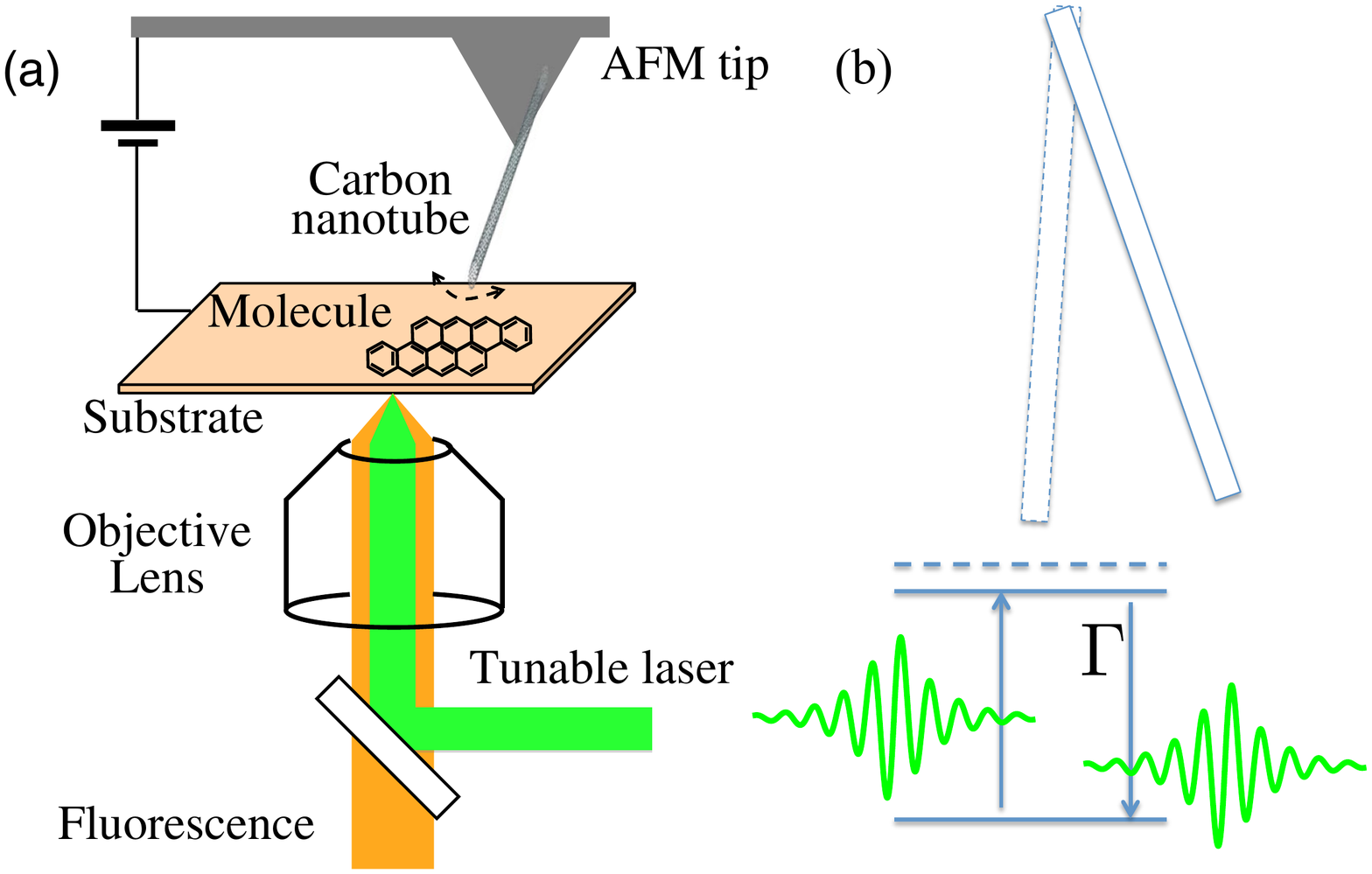}
  \caption{
  Left: Schematics of the system proposed in Ref.~\cite{puller_single_2013} and considered in this paper. Right: simplified view of the two-level system coupled to a mechanical oscillator via the Stark effect and to a laser.}
  \label{fig:Schematics}
\end{figure}

We consider a TLS coupled to a laser and to a mechanical oscillator as describled in 
Ref.~\onlinecite{puller_single_2013} (see Fig.~\ref{fig:Schematics}).
We will focus on this sytem, but the model describles several systems, 
for instance the TLSs coupled by the strain to an oscillator 
\cite{de_assis_strain-gradient_2017}. 
The coupling of the TSL to the mechanical oscillator is due to the Stark effect and the difference of 
potential that is induced between the oscillator (for instance a suspended carbon nanotube) and the 
transparent and conducting substrate over which the molecules are dispersed.
In experiments the light emitted by a single molecule is collected and detected as a function of the laser beam frequency and intensity. 
The system is describled by the following Hamiltonian:
\beq
	H_{\rm S}=-{\Delta\over 2} \sigma_z+\hbar \Omega \sigma_x \cos \omega_L t 
	- \hbar g \sigma_z (b+b^\dag) + \hbar \omega_m b^\dag b
	\,.
\eeq	
Here $\Delta$ is the TLS splitting, $\Omega$ the coupling intensity to the laser, 
$g$ the electromechanical coupling, $\omega_L$ is the laser frequency, 
and $\omega_m$ the mechanical pulsation ($\hbar$ is 
the reduced Planck constant). 
The operators $\sigma_x$ and $\sigma_z$ are Pauli matrices, 
and $b$ and $b^\dag$ are the destruction
and creation operators for the mechanical oscillator excitations.
The displacement operator reads $x=x_{z}(b+b^\dag)$ with 
$x_z=\sqrt{\hbar/2m\omega_m}$ the zero point displacement fluctuation amplitude
for an oscillator of mass $m$.
The TLS and the mechanical oscillator are coupled to the environnement that leads
to a finite linewidth $\Gamma$ of the TLS resonance and to a damping rate 
$\gamma$ for the oscillator.

\section{Born-Markov equations}
\label{sec:BornMarkov}

We proceed by assuming a weak coupling with the environment of the system. 
By standard methods \cite{schlosshauer_decoherence:_2007}  in the Born-Markov approximation the environnement can be traced out and an equation for the reduced density matrix 
$\rho(t)$ for the system (oscillator plus TLS) can be derived: 
\beq
	\dot \rho={\cal L} \rho .
	\label{BMEq}
\eeq
We define now $\rho_{ij}(x,x',t)=\langle x, i | \rho(t) | x', j\rangle$, where
$|x,i\rangle$ are the eigenstates of $x$ and $\sigma_z$ with eigenvalues 
$x$ and $(\sigma_z)_{ii}=\pm1$, respectively. 
It is convenient to introduce the variables $x_+=(x+x')/2$ and $x_-=x-x'$.
As discussed in the introduction we will consider in this paper the slow oscillator limit.
In terms of the above introduced parameters the condition reads
\beq
	\gamma \ll \omega_m, g \ll \Omega, \Gamma \ll \Delta \,.
\eeq
This implies a time scale separation between the TLS  and the mechanical oscillator dynamics.
It is thus convenient to write the Born-Markov operator 
as $\LL=\LLf+\LLs$ with $\LLf$ and $\LLs$ the fast and slow component, respectively.
Explicitly the fast component reads 
\beq
	\LLf
	=\left( 
\begin{array}{cccc}
0 & i \Omega/2 & - i \Omega/2 & \Gamma \\
i \Omega/2 & -i \delta'-\Gamma/2 & 0 &- i \Omega/2 \\
-i \Omega/2 & 0 &i \delta'-\Gamma/2 & i \Omega/2  \\
0 & -i \Omega/2 &  i \Omega/2 & -\Gamma \\
\end{array}
\right)
\eeq
with $\delta'=\delta-2 g x_+ /x_z$ and $\delta=\omega_L-\Delta/\hbar$, the detuning.
The components of the density matrix are $\{\rho_{11}, \rho_{12}, \rho_{21}, \rho_{22} \}$.
The slow component reads
\beq
  \LLs = {\cal L}_{osc} + {x_-\over x_z} \LL_-
\,,
\eeq
with
\beq
	{\cal L}_{\rm osc} 
	=
	{i \hbar \over m}  \partial_+ \partial_-
	-i {m \omega_m^2\over\hbar } x_+ x_- 
	-2\gamma x_-  \partial_-  
	-D x_-^2   
	\label{Losc}
\eeq
where $\LL_-= ig {\rm diag}(1,0,0,-1)$, and the last two terms in \refe{Losc} describe 
the coupling of the oscillator to an environment at temperature $T$. 
\notes{We use the notation of the Decoherence book, where $\ddot x=-2 \gamma \dot x$.}
From the fluctuation-dissipation theorem
$D=m \gamma (\hbar \omega_m/2) {\rm coth}(\hbar \omega_m/2 k_B T) $.
We use the notation $\partial_\pm = \partial/\partial x_{\pm}$.
\notes{$\partial_x=\partial_+/2+\partial_-$ and $ \partial_{x'}=\partial_+/2+\partial_-$}
The operator $\LLf$ describes the dynamics of the TLS, and 
implies a fast evolution of the density matrix on a scale of the maximum between
$\Gamma$ and $\Omega$.
Thus this evolution is much faster than that induced by  
the $\LLs$ term, that takes place on the $\omega_m$ time scale,
for what concern ${\cal L}_{\rm osc}$, and even slower for 
the dissipative part.
We have included the term proportional to $g x_+$ in $\LLf$
since we want to allow the possibility that this 
term becomes of the same order of $\Gamma$ in the strong 
coupling limit.
In order to check that the approximations are consistent 
what matters is the fluctuation of the oscillator position 
$(\Delta x_+)^2=\langle x_+^2\rangle  - \langle x_+\rangle^2$. 
If  $g \Delta x_+/x_z $ is of the order or larger than $\omega_m$, 
then this coupling has to be included in the fast part.  
On the other side, it is typically correct to regard  
the term proportional to $g  x_- $ as a slow contribution that can
be included in $\LLs$. 
This second assumption is valid for $g \Delta x_-/x_z  \ll \Gamma$.
We will come back to these two assumptions after the solution of the  
\refe{BMEq} has been obtained.

\section{Adiabatic elimination of fast variables}
\label{sec:adiabatic}

We now exploit the separation of time scales, $\omega_m \ll \Gamma$,
using the method of adiabatic elimination of fast variable (see for instance 
\cite{gardiner_stochastic_2009}) to integrate out the TLS fast degrees of freedom and obtain an 
equation for the mechanical oscillator reduced density matrix.
We begin by noting that $\LLf$ is a function of $x_+$ only.
One can then define the kernel of the operator $\LLf$ by the equation:
\beq
   \LLf(x_+) \rho^{0}(x_+)=0
	\,.
\eeq
Physically $\rho^0(x_+)$ is the stationary state of the TLS for given value of $x_+$.
Since $\LLf$ is not Hermitean left and right eigenvectors are different. 
Let's define $w_0$ as the left eigenvector of $\LLf$ with vanishing eigenvalue.
One can readily show that it has the form $w_0=\{1,0,0,1\}$ and when projected on an arbitrary state
$\rho$ it gives its trace over the TLS states: 
\beq
	(w_0,\rho)=\rho_{11}+\rho_{22}={\rm Tr} \rho
		\,.
\eeq
Since probability is conserved by the time evolution 
$0=d{\rm Tr}(\rho)/dt= {\rm Tr}(\LLf \rho)=(w_0,\LLf \rho)$ for any $\rho$, the relation
$w^t_0 \LLf=0$ holds. 
We choose now the normalization of $\rho^{0}(x_+)$ such that $(w_0,\rho^{0})=1$.
Any $\rho$ can then be written as a sum of its projection on the kernel 
of $\LLf$, the slow component, and on its othogonal complement, the fast component:
\beq
	\rho_{ij}(x_+,x-,t)=\rho_{ij}^{0}(x_+) R(x_+,x_-,t)+ \rho^{f}_{ij}(x_+,x_-,t)
	,
	\label{vector}
\eeq
where by construction $(w_0,\rho^f)=0$ and $(w_0,\rho)=R(x_+,x_-)$. 
Let us define $P$ as the projector on $\rho^0$, and $Q=1-P$ as the projector on the complement.
Note that $P$ and $Q$ depend on $x_+$, they are the projectors on the 4$\times$4  
phase space of the TLS.
Again for given $x_+$, the following properties hold : $P \LLf = \LLf P=0$.
We substitute now this expression in the master \refe{BMEq} and by applying 
alternatively $P$ and $Q$ we obtain the two equations:
\beqa
	\dot R \rho^0 &=& P \LLs (R \rho^0+\rho^f) ,
		\label{sys1}
		\\
	Q \dot {\rho^s} &=&Q \LLf  \rho^f+Q \LLs	(R \rho^0+\rho^f)
		\,.
		\label{sys2}
\eeqa
Up to now these equations are exact. Let's now use the fact that $\LL_f$ is very large to solve the 
second equation (a systematic expansion could be derived using the Laplace transform, we 
consider here only the leading term)
\beq
	 Q \LLf \rho^f=-Q \LLs (R \rho^0)
	 \,.
\eeq
Since $\LLf $ acts on the subspace orthogonal to the kernel and 
the solution is seeked in the same space $Q \LLf Q$ is actually invertible in this subspace 
(while $\LLf$ has one vanishing eigenvalue in the full space of $\rho$).
We thus substitute the solution into \refe{sys1} and obtain:
\beq
	\dot R \rho^0 = P \LL_{\rm s} (R \rho^0)	-P\LL_s (Q\LL_fQ)^{-1} Q \LL_s (R \rho^0) 
	\label{Master2}
		.
\eeq
We begin by evaluating 
\beq
		Q\LL_s (R \rho^0) = {x_-\over x_z} R Q \LL_- \rho^0 
		+{i \hbar \over m}  \partial_-R Q \partial_+ \rho^0
		.
\eeq
By projecting \refe{Master2} on $w_0$ we obtain
\beq
		\dot R = 
		\left[
			{\cal L}_{\rm osc} 
			+ {x_-\over x_z} \alpha_1 
			-{i \hbar x_- \over m x_z} \alpha_2  \partial_-
			+  {x^2_- \over x^2_z} \alpha_3 
		\right] R
		\label{dotR}
\eeq
where 
$\alpha_1 = (w_0, \LL_- \rho^0)$, 
$\alpha_2 = (w_0, \LL_- (Q \LLf Q)^{-1} Q \partial_+ \rho^0  )$, and
$\alpha_3 = (w_0, \LL_- (Q\LL_fQ)^{-1} Q \LL_- \rho^0)$.
	
Using the explicit form of $\LLf$ one can readily find the three matrix elements:
\beqa
	\beta_1 &=& -{i \alpha_1 \over g}=\frac{\Gamma ^2+4 {\delta'} ^2}{\Gamma ^2+4 {\delta'}^2+2 \Omega ^2} ,
	\\
	\beta_2 &=& {i x_z\Gamma^2 \alpha_2 \over 2 g^2}
	=
		-\frac{32 {\delta'}  \Gamma \Omega ^2 \left(2 \Gamma ^2+\Omega ^2\right)}{ \left(\Gamma ^2+
		4 {\delta'}^2+2 		\Omega ^2\right)^3}  ,	\\
	\beta_3 &=&
	 {\alpha_3 \Gamma \over g^2}  
	 	=
		\frac{4 \Omega^2 \left(\Gamma ^2+4 {\delta'} ^2\right) \left(2\Gamma ^2+\Omega ^2\right)}
		  {\left(
		  \Gamma ^2+4 {\delta'}^2+2 \Omega^2
		  \right)^3}  .
\eeqa
Substituting these expressions into \refe{dotR} and introducing the 
Wigner transform
 $W(x_+,p)=\int dx_- e^{-i p x_-} R(x_+,x_-)$ we have $x_-\rightarrow i \hbar \partial_p$ and $\partial_-\rightarrow ip/\hbar$ in \refe{Master2}. 
We thus find for $W$ the following equation:
\beq
	\dot W
	=
	\left[  -{p\over m} \partial_{+} + (m\omega_m^2 x_+ - F)   \partial_p
    + 2\gamma_t \partial_p p + D_t \partial_p^2 \right] W
	\label{FPEq}
\eeq
where $F$ is the average force acting on the oscillator:
\beq
	 F = { \hbar g \over x_z} \langle \sigma_z\rangle =
	- {g \beta_1 \over x_z}
	\label{force}
	,
\eeq
$\gamma_t =\gamma+\gamma_o$ and 
$D_t=D+D_o$, are the total dissipation and diffusion coefficients, respectively,
with
$\gamma_o=\hbar g^2 \beta_2/mx_z^2 \Gamma^2$ 
and 
$D_o = \hbar^2 g^2 \beta_3/x_z^2 \Gamma$
the dissipation and the diffusion coefficients 
induced by the coupling to the driven TLS. 

Equation (\ref{FPEq}) with the relations (\ref{sys2}) and (\ref{vector}),
allow to obtain the behaviour of the mechanical oscillator and the 
optical response of the TLS.
In practice $\rho^f$ is very important for the derivation of the equation of motion 
of $R$, but its contribution to $\rho$ is small (of the order $\omega_m/\Gamma$), 
and it can be neglected in the calculation of the averages in the following.

\section{Effective temperature}

\label{sec:EffTemp}

\newcommand{\Teff}{T_{\rm eff}}

We begin the study of \refe{FPEq} by defining an effective temperature in analogy with the 
fluctuation-dissipation relation:
\beq
	\coth \left( \hbar \omega_m \over 2 k_B \Teff \right)={2D_t \over \gamma_t \hbar \omega_m}
	\,.
\eeq
In the case of $\gamma \ll \gamma_0$ one finds 
\beq
	\coth \left( \hbar \omega_m \over 2 k_B \Teff \right) 
	= 
	- {\Gamma^2+{\delta'}^2 \over 4 {\delta'} \omega_m}
	\label{coth}
	\,.
\eeq
For positive values of $\delta'$ the system is unstable (negative damping term). 
For negative values of $\delta'$ the function in \refe{coth} has 
a minimum value of $\Gamma/4\omega_m\gg1$ at $\delta'=-\Gamma$.
This means that when the coupling to the TLS dominates over the coupling to the environment, the system reaches a classical stationary state ($k_B \Teff \gg \hbar \omega_m$). 
Thus for $\gamma=0$ one can write 
\beq
	k_B \Teff
	=
	\hbar {\Gamma^2+{\delta'}^2 \over 8 {\delta'}}
	\,.
	\label{Teff}
 \eeq
We note that the form of the effective temperature is the same that is found for 
an oscillator coupled to a cavity \cite{marquardt_quantum_2007}.
This follows for the similarity of the spectrum of fluctuation of $\sigma_z$ and that 
of the number of photons in a cavity. 
The result implies clearly that it is not possible to use the TLS in the 
$\Gamma\gg \omega_m$ limit to cool the oscillator in the quantum regime \cite{puller_single_2013,pistolesi_cooling_2009}.

We are now in the position to check the conditions
on $\Delta x_+$ and $\Delta x_-$ assumed for the solution 
of the problem.
From the equipartition theorem 
$m \omega_m^2 (\Delta x_+)^2 = k_B \Teff$ 
and $\Delta x_-\sim \hbar /\Delta p $  with $(\Delta p)^2/m=k_B \Teff$.
The condition on $\Delta x_-$ reads then
\beq
	g \ll \Gamma \left(k_B T_{\rm eff} \over \hbar \omega_m\right)^{1/2}
		\label{condition}
		\,.
\eeq
%
%
Since $k_B T_{\rm eff} > \hbar \Gamma$, a sufficient condition 
for the validity of the approximations is $g \ll \Gamma (\Gamma/\omega_m)^{1/2}$, 
that gives a large window of validity of the theory.

For finite value of $\gamma$ a part of the region for which $\delta'>0$ 
becomes stable, but before discussing this point we need to take into account the fact that $\delta'=\delta-2g x_+ /x_z$ and that this induces a mechanical bistability.

\section{Mechanical bistability}

\label{sec:mechanicalbistability}

The value of $x_+$ entering $\delta'$ is a stochastic variable whose statistics can be 
obtained from the solution of \refe{FPEq}. 
In general the distribution function is strongly peaked around the equilibrium position
$x_e$ that satisfies the equation
\beq
	m \omega_m^2 x_e=F(x_e) 
	\label{eqEq}
	\,.
\eeq
We thus begin by solving \refe{eqEq}.
The equation can be studied more conveniently by eliminating 
\beq
	x_e=(\delta-\delta') x_z / 2g
	\label{eqXe}
\eeq
from the definition of $\delta'(x_e)$. 
The equation then reads
\beq
	\delta-\delta' 
	= 
	\lambda \Gamma \beta_1(\delta')
	\label{InstEq}
\,,
\eeq
where we defined $\lambda=4g^2/\omega_m\Gamma$ that 
is the relevant dimensionless coupling constant also known 
as cooperativity. 

A relevant interpretation of $\lambda$ for our problem is also the following: 
When the TLS is in the excited state an additional force $F_o=\hbar g/x_z$ acts 
on the oscillator modifying its equilibrium position 
$\Delta x_e= F_0/m \omega_m^2 $.
Let us call  $\epsilon_P=F_o \Delta x_e=\hbar g^2/\omega_m$
the (classical) energy scale that corresponds to (twice) the variation 
of the potential energy of the oscillator.
Comparing it with the relevant energy scale of the TLS $\hbar \Gamma$ we have 
$\epsilon_P/\hbar \Gamma=\lambda/2$, that measures the relevance of the 
TLS on the oscillator dynamics.
This phenomenon resembles to the bistability expected in suspeded carbon nanotubes
forming a single-electron transistor \cite{pistolesi_self-consistent_2008,micchi_mechanical_2015}.
There the two-level system is the empy or filled suspended quantum dot, and the role of the laser driving is played by the electrons entering the quantum dot for transport. 
We come now back to \refe{InstEq}. 
Comparing the derivatives with respect to $\delta'$ 
of the right- and left-hand sides of \refe{InstEq} one finds that 
for
$
\lambda  <  \lambda_c = 2(\Gamma^2+2\Omega^2)^{3/2}/\Omega^2 3\sqrt{3}
$
there is a single solution for the equation for any value of $\delta$.
For $\lambda>\lambda_c$ three solutions exist, 
two stable and one unstable.
The two stable solutions for $\delta'$ correspond to two stable (or metastable) 
equilibrium positions given by \refe{eqXe}.
From the expression of $\lambda_c$ one can see that $\lambda_c$ takes the 
minimum value of 2 for $\Gamma=\Omega$. 
In terms of the bare coupling $g$ this implies that the minimum value required to observe the bistability is 
$g_{\rm min}=(\omega_m \Gamma/2)^{1/2}$, 
the geometric 
mean of the mechanical frequency and of the TLS inverse lifetime.
Equivalently, the requirement is that the cooperativity $\lambda>1$.

\begin{figure}[tbp]
  \includegraphics[width=3in]{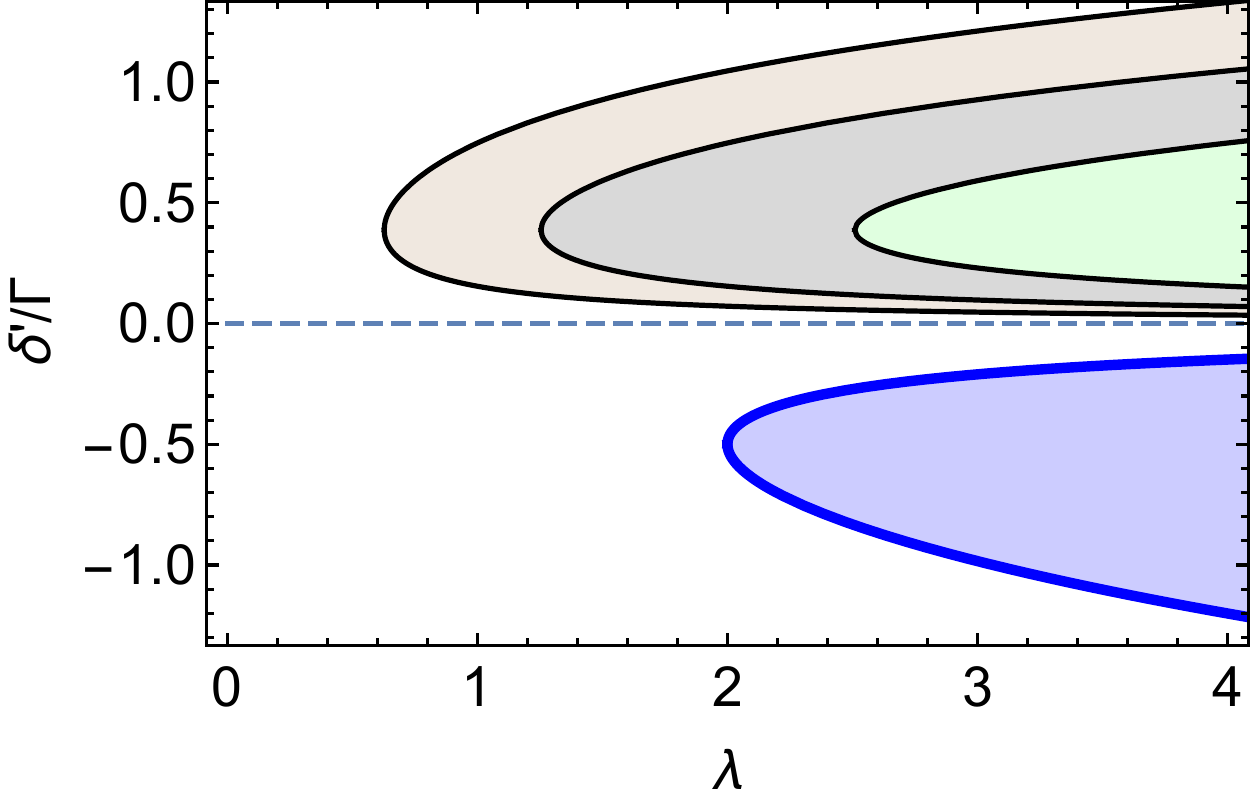}
  \caption{
  Region of static (shaded blue region for negative $\delta'$) 
  and dynamical (shaded regions for positive $\delta'$) instabilities 
  in the plane $\lambda$-$\delta'$ for $\Omega=\Gamma$ and for 
  $Q \omega_m/\Gamma=$4 (left most brown region), 2 (grey), 1 (green).
The dashed line corresponds to the value $\delta'=0$, 
separatrix between cooling and heating in the case $\gamma=0$ (or $Q=\infty$).
}
\label{fig:phasedeltaP}
\end{figure}

\begin{figure}[tbp]
  \includegraphics[width=3in]{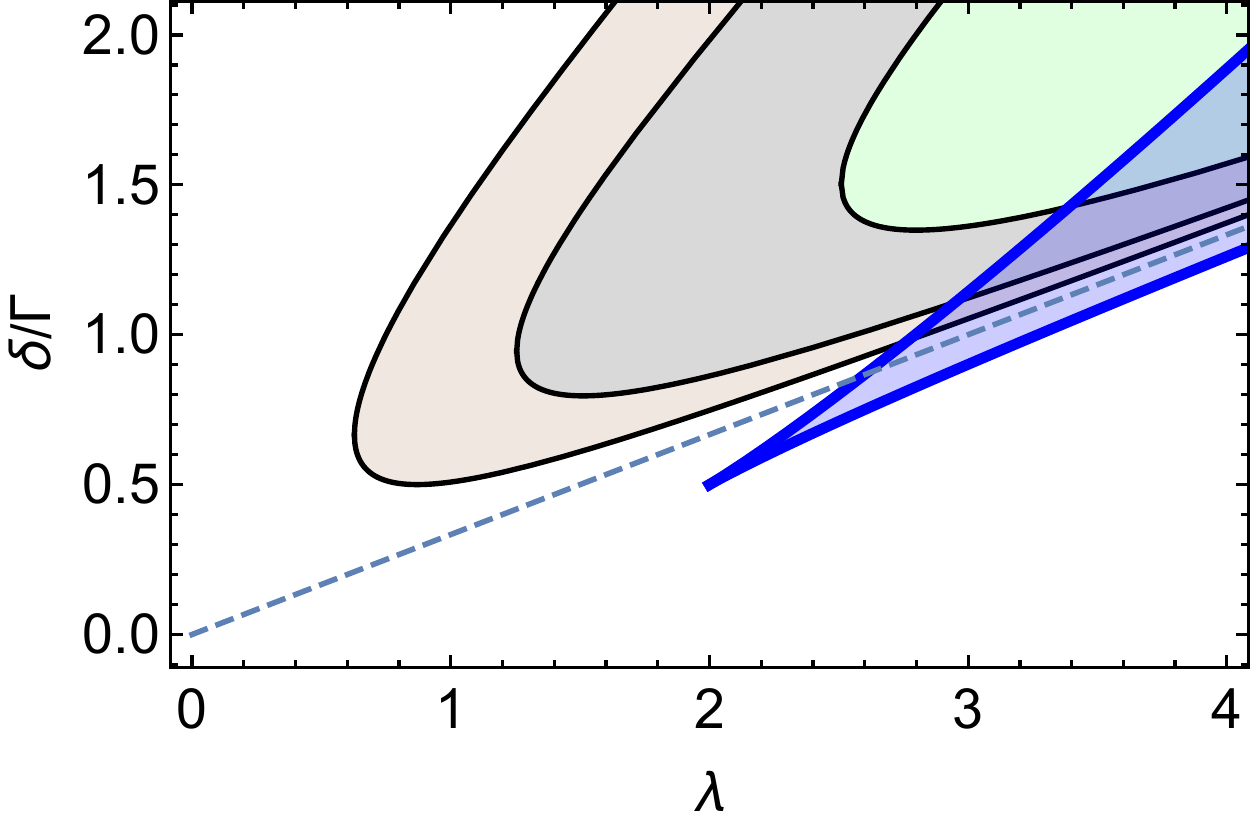}
  \caption{
Same as Fig.~\ref{fig:phasedeltaP} in the plane $\delta$-$\lambda$.
Due to the bistability the contour of the bistability region can be in coincidence 
with the dynamically unstable region of the phase space.
 }
  \label{fig:phasedelta}
\end{figure}

In order to find the region of bistability as a function of $\lambda$ and the 
physical laser detuning $\delta$, we begin by recognizing that the bistable behaviour 
upon increasing $\lambda$ begins when the $d/d\delta'$ of the left and right-hand side of \refe{InstEq} coincide:
\beq
	-1=\lambda \Gamma {d \beta_1 \over d \delta'} .
\eeq
This is an equation for $\delta'$ and $\lambda$.
Its solution is shown in Fig.~\ref{fig:phasedeltaP} for $\Omega=\Gamma$.
One can see that the bistable region takes place for negative values of $\delta'$, for which the TLS generate a standard positive dissipative term. 
One should however be cautious when converting this result to the externally tunable 
$\delta$. 
This can be done by using \refe{InstEq}.
The result is shown in Fig.~\ref{fig:phasedelta}
The dashed line is the region $\delta'=0$. As one can see a part of the bistable phase 
is now apparently in the region of dynamical instability ($\delta'>0$). 
This is more subtle, since actually the transformation (\ref{InstEq}) 
is not bijective, to the same value of $\delta$ two values of $\delta'$ 
may be associated.
We will discuss later the consequences of this fact in more details.

\begin{figure}[tbp]
  \includegraphics[width=3in]{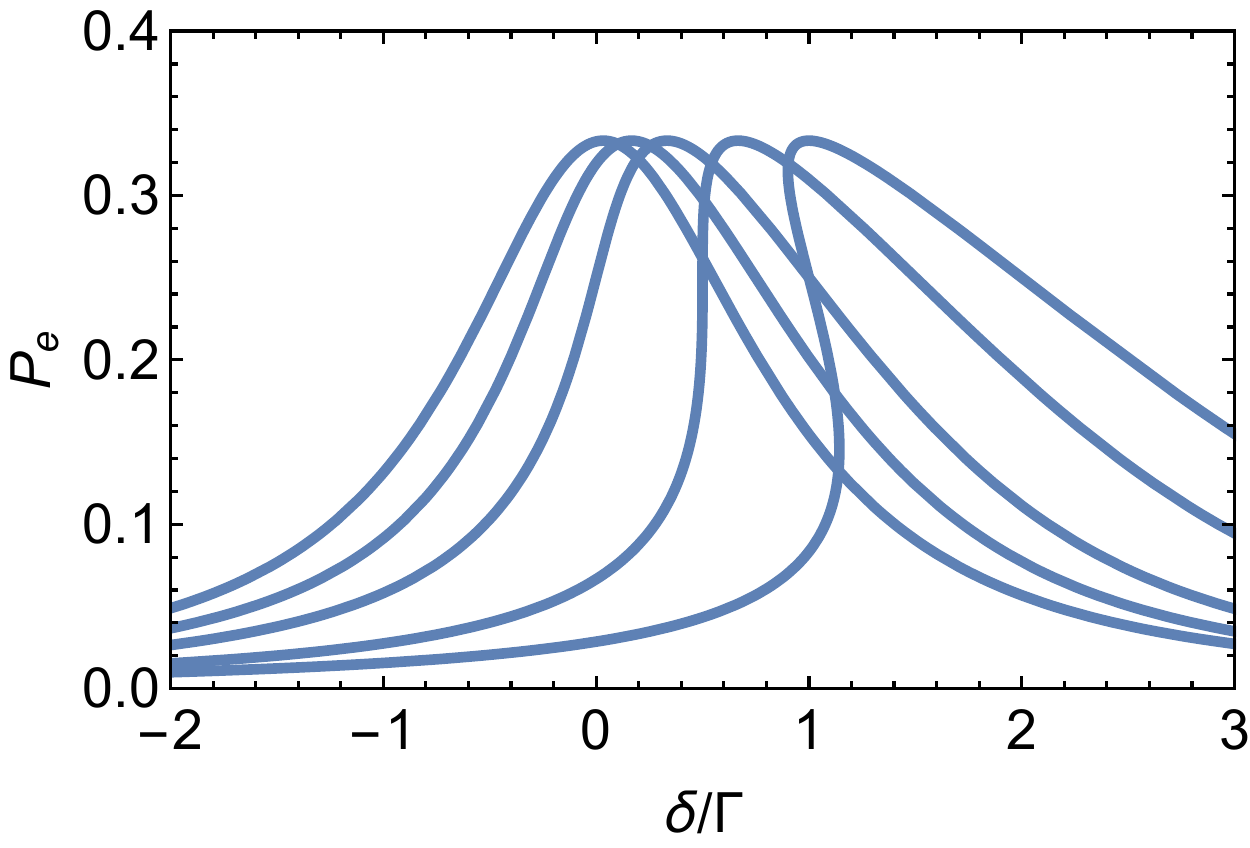}
  \caption{
 Probability of occupation of the excited state 
 ($P_e$ proportional to the luminescence) as a function of 
  the detuning $\delta$ for   $\Omega=\Gamma$ and 
  $\lambda=0.1, 0.5, 1, 2$ and 3 (from left to right). 
  The bistable behavior begins at $\lambda=2$.
 }
  \label{fig:Lumin_MF}
\end{figure}

Let us begin by studying the consequences of the bistability on the luminescence.
In order to obtain the luminescence we only need the probability of occupation of the 
excited state, $P_{\rm e}$, as a function of the externally fixed detuning $\delta$.
From the solution of $\LLf \rho^0=0$ one finds the familiar result:
\beq
	P_{\rm e} = {\Omega^2 \over \Gamma^2+2 \Omega^2+4 {\delta'}^2}
	\,.
\eeq
We plot parametrically $P_e$ and $\delta(\delta')$ from \refe{InstEq} in 
Fig.~\ref{fig:Lumin_MF}. 
One can see that the lineshape is strongly modified, with the apperance of a part 
with three values possibles, corresponding to the two stable and one unstable 
equilibrium state. 
Clearly the unstable state cannot be realized, but the oscillator spends 
a sizable part of the time on the two (meta-)stable states.
In order to evaluate the actual form of the expected luminescence line we need to 
solve the Fokker-Planck \refe{FPEq}.

\section{Effect of fluctuations}
\label{sec:stochastic}

We need to consider the effect of fluctuations.
Since the system may be unstable for positive values of $\delta'$, one cannot discard anymore the intrinsic dissipation due to the coupling to the 
environnement at temperature $T$.
The region of dynamical instability is defined by the condition
$\gamma_t < 0$, where $\gamma_t$ is defined after \refe{force}.
This gives the equation for the critical line:
\beq
	1+2 \lambda Q  {\omega_m \over \Gamma} 
	\beta_2(\delta')=0
	\label{criticalGammat}
	\,,
\eeq
where we introduced the quality factor $Q=\omega_m/\gamma$. 
We show in Figs. \ref{fig:phasedeltaP} and \ref{fig:phasedelta} 
the regions of dynamical instability as a function of $\lambda$ and $\delta'$ or $\delta$, respectivity.
It is important to realize that the dynamical instability takes place very close to the 
static bistability, we will see that this has consequences in the expected 
lineshapes.

Let us now discuss the interplay of the heating/cooling effect and the bistable behaviour.
\begin{figure}[tbp]
  \includegraphics[width=1.5in]{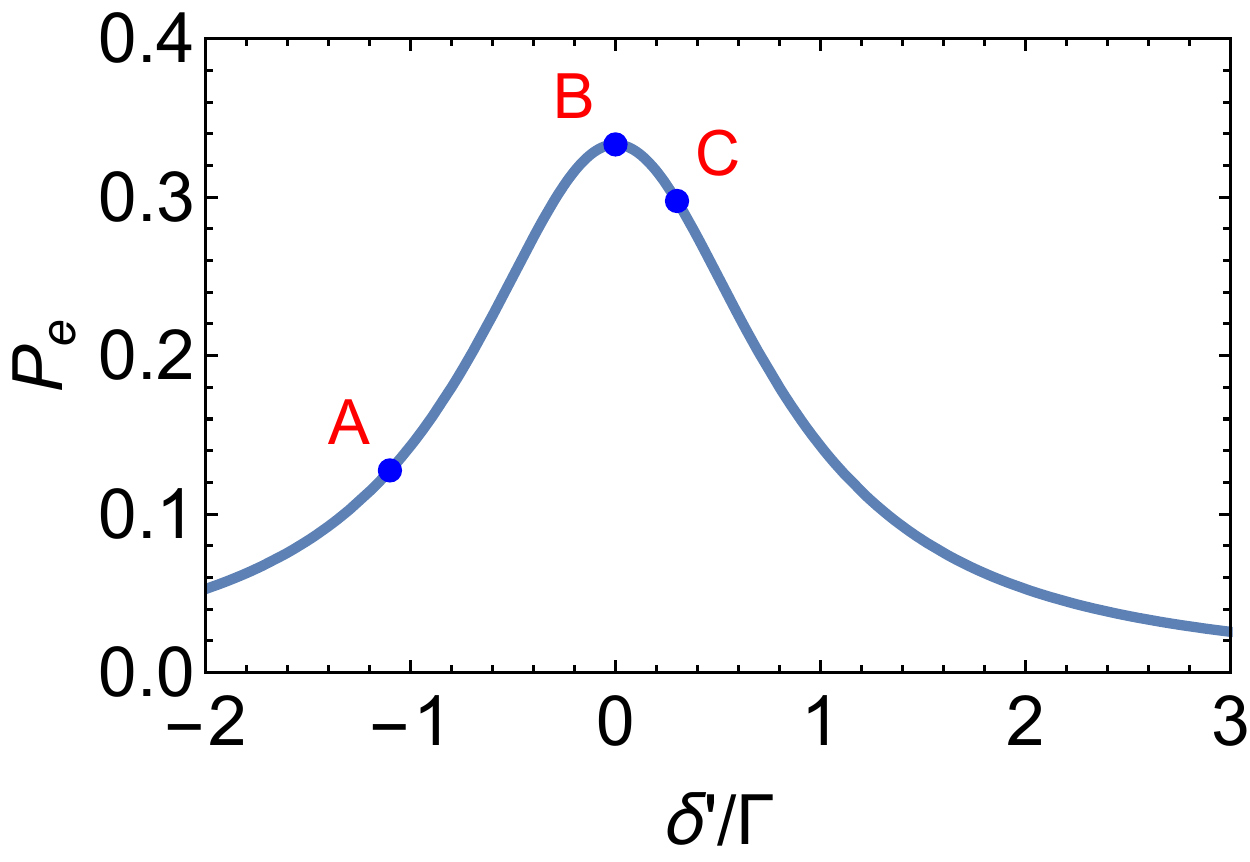}
  \includegraphics[width=1.5in]{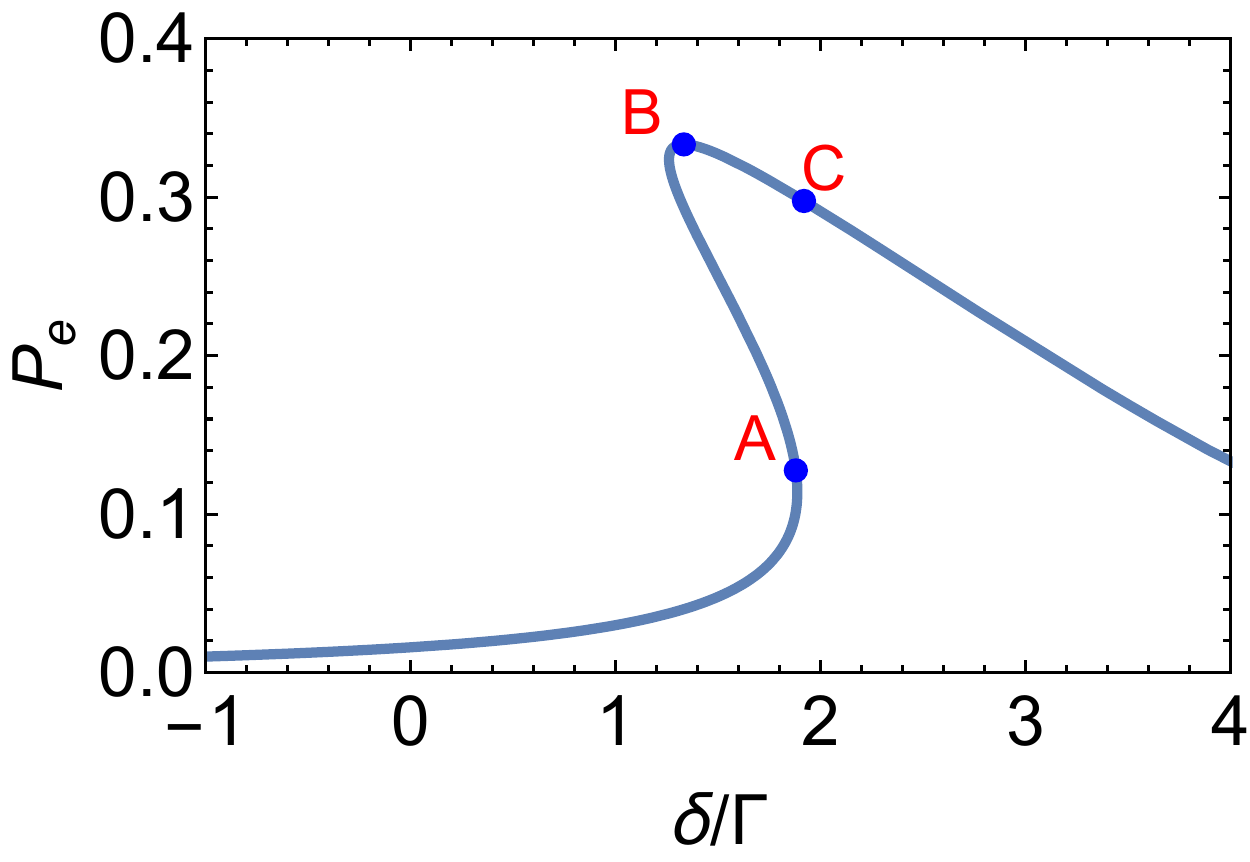}
  \caption{
Mapping of the luminescence line from $\delta'$ to $\delta$. The points 
$A$, $B$, and $C$, correspond to cooling, neutral, and heating values, respectively.
One can see that for $\delta=2$ both cooling and heating are possible,
depending on value of $x$ that determines if the point $A$ or $C$ is actually occupied. 
}
\label{fig:LuminABC}
\end{figure}
In Fig.~\ref{fig:LuminABC} we show $P_{\rm e}$ as a function of $\delta'$ and 
$\delta$.
One can follow how the states indicated by the letters $A$, $B$, and $C$ 
are mapped in the $\delta$ plot.
The point $A$ is in the cooling part of the line ($\delta'<0$) but in the 
$\delta$ space it appears at value of $\delta$ larger than the value of 
$\delta$ corresponding to $B$ ($\delta'=0$), that defines 
the border between cooling and heating.
The cooling state $A$ is thus bistable with the heating state $C$.
Due to the fluctuations the system will spend some time in both states,
with a probability that is determined by Fokker-Planck equation. 
This also explains the shape of the static bistability region, that apparently leaks on
the dynamically unstable region, as shown in Fig.~\ref{fig:phasedelta}.

In order to find the contribution of the fluctuations quantitatively we solve numerically
the Fokker-Planck \refe{FPEq} by discretizing the phase space $x$-$p$ 
in $N_x$ and $N_p$ points, respectively.
The operator entering \refe{FPEq} and acting on $W$ becomes thus a matrix of dimension $N=N_x N_p$.
The stationary solution of \refe{FPEq} is found by solving the equation with the 
constraint of the normalization of $W$.
We find that typically $N_x = N_y = 100$ is already sufficient to obtain a solution 
of the equation in the range of interest of the parameters. 
The Fokker-Planck equation can be rewritten in terms of the dimensionless variables
$\tilde x=x/\Delta x_e$, $\tilde p = p \omega_m/F_o$, and
$\tilde t=\omega_m t$. 
This gives
\beq
	{\partial W \over \partial \tilde t} 
	=
	\left[  
	- \tilde p \partial_{\tilde x} 
	+ [\tilde x  -\beta_1]  {\partial_{\tilde p} }
	+ \tilde \gamma_t \partial_{\tilde p} {\tilde p} 
	+ \tilde D_t \partial_{\tilde p}^2 \right] 
	W
\eeq
where
\beq
	\tilde \gamma_t= {\omega_m\over \Gamma}
	\left[{\lambda \beta_2\over 2}+ {\Gamma \over Q \omega_m} \right]
	,\qquad
	\tilde D_t=-{\omega_m\over \Gamma}\beta_3
	+{k_B T \over \epsilon_P Q}
	.
\eeq

\begin{figure}[tbp]
  \includegraphics[width=3in]{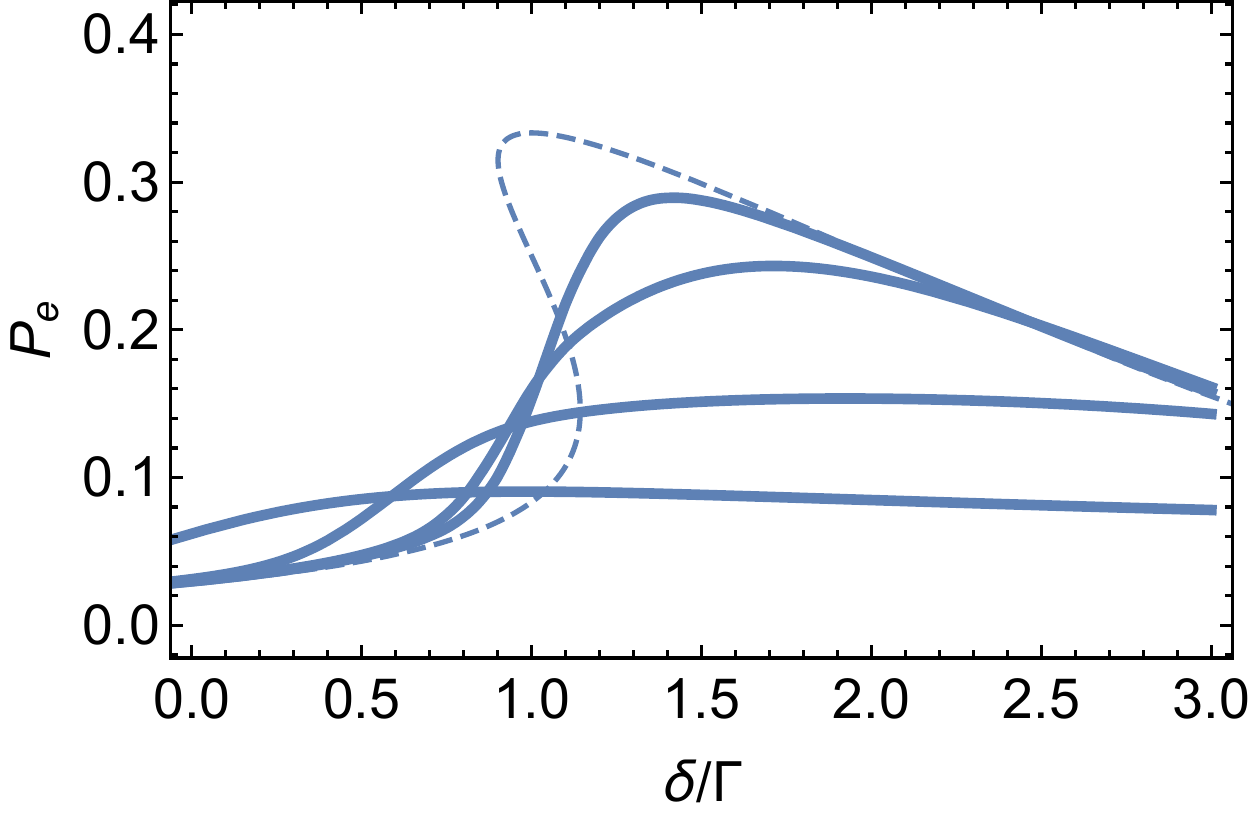}
  \caption{
  Luminescence:
  $\lambda=2$, $\omega_m/\Gamma=10^{-3}$, $k_B T/\epsilon_P=0.01$, $Q=$10, $10^2$, $10^3$, and $10^4$ for the curves from the stepest to the smoothest, respectively. 
Increasing $Q$ increases the fluctuations and smothen the lineshape.
The parameter $Q \omega_m/\Gamma$ entering \refe{criticalGammat} 
takes thus the values $10^{-2}$, $10^{-1}$, $1$, and $10$.
  } 
  \label{fig:Lumin_FP}
\end{figure}

Results of the numerical solutions for the luminescence are shown in Fig.~\ref{fig:Lumin_FP} 
for $k_B T/\epsilon_P=0.01$, $\omega_m/\Gamma=10^{-3}$, 
$\lambda=2$ and for different values of $Q$. 
For small $Q$ the luminescence follows closely the mean field result (shown dashed) apart for the bistable region.
Increasing $Q$ one enters the region of dynamical instability, as expected from 
\refe{criticalGammat} and Fig.~\ref{fig:phasedelta}, and fluctuations increase dramatically, with the system spending a sizable time on the heating region.
\begin{figure}[tbp]
  \includegraphics[width=3in]{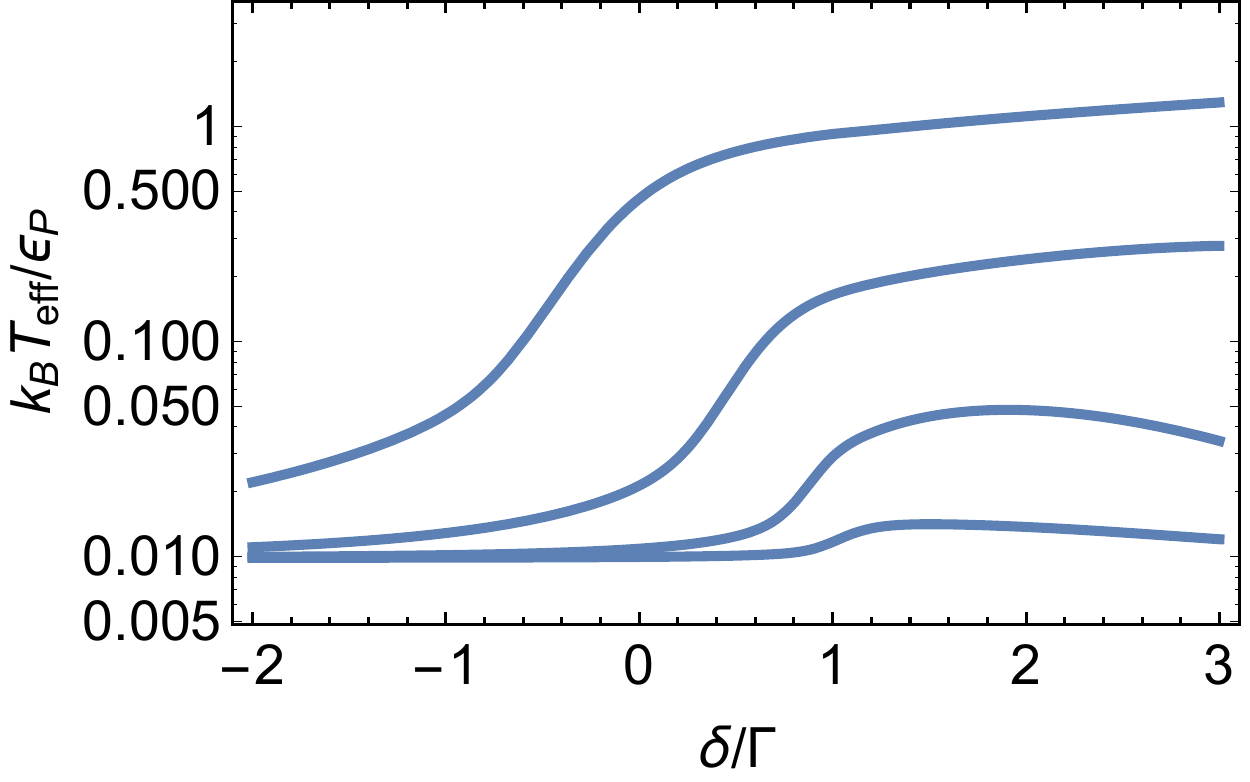}
  \caption{
  Effective temperature $k_B T_{\rm eff}/\epsilon_P$ as a function of the detuning $\delta$ for the same value of the quality factors of Fig. \ref{fig:Lumin_MF}. 
  The higher values of the temperature are of course obtained for the highest $Q$. 
  }
  \label{fig:EffectiveTemp}
\end{figure}
This is confirmed by the dependence of the effective temperature on $\delta$
for given value of $Q$, as shown in Fig.~\ref{fig:EffectiveTemp}.
One can note that large values of $Q$ the effective temperature 
increase for positive values of $\delta$ reaching values of the order of 
$\epsilon_P$, and thus washing out the bistable behavior. 
The result is a very smooth and asymmetric lineshape.

\section{Conclusions}
\label{sec:conclusions}

In this paper we studied the behavior of a slow mechanical resonator 
($\Gamma\gg \omega_m$) coupled to a laser driven TLS in the strong coupling regime. 
We began by the Born-Markov description of the system 
[cf. \refe{BMEq}], and then, by eliminating the TLS fast variables we obtained
\refe{FPEq}] for the Wigner function of the mechanical oscillator.

By analyzing this equation we showed that the oscillator effective temperature can be 
controlled by the laser detuning and the coupling (cooling or heating).
We found that, when the 
coupling of the TLS dominates over the coupling to the environment, the mechanical oscillator 
can only be in a classical regime ($k_B T_{\rm eff} \gg \hbar \omega_m$).

We showed than that for sufficiently strong coupling the mechanical system can undergo a bistability. The conditions on the cooperativity and detuning for its observation are resumed in
Fig.~\ref{fig:phasedelta}.
Contrary to optomechanical cavities, the coupling controlling the instability cannot be tuned by the laser intensity (that here is parametrized by $\Omega$). 
As a rule of the thumb, we find that a coupling constant of the order of $(\omega_m \Gamma)^{1/2}$ is necessary for the bistability to take place. 
The bistability allows the possibility that the two stable states are one in the cooling and the other in the heating regime. This leads to a peculiar shape of the luminescence linewidth
(cf.~Fig.~\ref{fig:Lumin_FP}) , that is broaden mainly due to the increase of the oscillator fluctuations induced by the heating effect. 
A promising experimental system where this effect could be observed is the one
proposed in Ref.~\cite{puller_single_2013} of single molecules coupled to a carbon nanotubes. 
The typical values of the parameters are 
$m=10^{-21}$ Kg,
$\Gamma > 10^7$ Hz,
$g < 10^9$ Hz, with a mechanical frequency of the carbon nanotube that 
depending on its length can vary from kHz to MHz. 
With these numbers it should be possible to reach large values of $\lambda$, 
for instance for $g=10^6$ Hz, $\omega_m=2\pi \cdot 10^{5}$ Hz, $\Gamma=10^7 $ Hz
one finds $\lambda \approx 0.5 $. 
For these couplings the effective mechanical quality factor $Q_o=\omega_m/\gamma_o$ induced by the coupling to the oscillator is $(\Gamma/g)^2 \beta_2/2 \approx 10^2$.
Thus even if the scale of the energy barrier between the two stable states ($\epsilon_P$)
is very small, when the coupling to the environnement is sufficiently weak ($Q=\omega_m/\gamma\gg Q_o$) the effective temperature is controlled 
uniquely by the TLS and $k_B T_{\rm eff} \ll \epsilon_P$.
This should lead to the observation of a luminescence line similar to the one predicted in 
Fig.~\ref{fig:Lumin_FP}.

Concerning the approximations used in the paper, it turns out that one of the main condition  
is that the spread of the quantum variable $x_-$ is small: $\Delta x_-/x_z \ll  \Gamma/g$.
This is necessary to include the term $g x_- $ in the slow component ($\LLs$) of the master equation, leaving the fast operator independent on $x_-$. 
A less technical way of stating this condition is to say that it 
fixes a limit on the quantum nature of the mechanical degree of freedom. 
In terms of the coupling constant the condition is given by \refe{condition}.
One finds that the scale of validity is set by $g \ll \Gamma (\Gamma/\omega_m)^{1/2}$
 [or $\lambda \ll  (\Gamma/\omega_m)^2$]. 
The bistable behavior as describled above is thus well inside the limit of validity of the  approximation.
For $g \gg (\Gamma/\omega_m)^{1/2}$ the coupling is so strong, that quantum coherence cannot be neglected anymore between two transition events, even if $\Gamma \gg \omega_m$. 
This regime if of course relevant for investigatigations on quantum manipulation of the mechanical states and constitutes an interesting perspective for future work.

\begin{acknowledgements}
I thank the {\em Conseil r\'egional de Nouvelle-Aquitaine} for financial support.
I also thank V. Puller, R. Avriller, and C. Dutreix, for discussions. 
\end{acknowledgements}

\appendix

\section{Relation to the quantum noise method} 

Let us briefly comment on the physical interpretation of the fluctuation and dissipation terms.
For weak coupling it is known \cite{clerk_quantum-limited_2004} that the dissipation and 
fluctuation can be derived directly from the quantum correlation function of the force operator. 
In our case this is defined as: 
$ S_{FF}(t) = (\hbar g/x_z)^2 S_{zz}(t)$ with 
\beq
	S_{zz}(t)=\langle \sigma_z(t) \sigma_z(0)\rangle - \langle \sigma_z\rangle^2 ,
\eeq
where the time evolution is ruled by only the TLS part of the Hamiltonian ($g=0$).
The quantity $S_{zz}$ can be obtained explicitly:
\beq
	S_{zz}(t>0)=(w_0, M_z e^{\LLf t} M_z \rho^0)-(w_0,M_z \rho^0)	^2
\eeq
where $M_z=(\LL_- + \LL_+)/(ig)$ is the superoperator for $\sigma_z$ and 
we introduced $\LL_+=ig {\rm diag}(0,1,-1,0)$ by analogy with the definition of $\LL_-$.
Introducing the Laplace transform 
$S_{zz}(s)=\int_0^{+\infty} dt e^{st} S_{zz}(t)$  
(with ${\rm Re} s<0$)
we have 
\beq
	S_{zz}(s)=(w_0,(M_z)\rho^0)^2/s - (w_0,M_z(s-\LLf)^{-1}M_z \rho^0)
	\,.
\eeq
Using the projectors $Q$ and $P$ one can readily show that 
\beq
	S_{zz}(s) = - (w_0, M_z Q (s-Q\LLf Q)^{-1} Q M_z \rho^0) .
	\label{eqSzz}
\eeq
For $s=0$ the inverse has to be performed in the sub-space 
defined by the projector $Q$.
The power spectrum 
$S_{zz}(\omega)=\int_{-\infty}^{+\infty} dt e^{i\omega t} S_{zz}(t)$ can then be related 
directly to the Laplace transform by using the property $S_{zz}^*(t)=S_{zz}(-t)$ 
that leads to 
\beq
	S_{zz}(\omega)=2 {\rm Re}[S_{zz}(s=i\omega-0^+)].
\eeq
We note that $w_0 \LL_+=0$. By explicit calculation one can verify that the 
term proportianal to $\LL_+$ coming from the $M_z$ on the right matrix element of 
\refe{eqSzz} gives an imaginary term for $s=0$, and thus it does not contribute to 
$S_{zz}(\omega\rightarrow 0)$.
One can thus substitute $\LL_-$ into the definition of $M_z=\LL_-/(ig)$ in this case. 
For $\omega_m\ll \Gamma$ we can obtain 
the diffusion constant of the Fokker-Plack equation from the vanishing frequency value of
the force fluctuation spectrum:
\beq
	D_0=S_{FF}(\omega\rightarrow0)/2 = \hbar^2\alpha_3/x_z^2 .
\eeq
This value coincides with the adiabatic approach result.
One can also verifiy by direct calculation that the derivative with respect to $\omega$ 
of $S_{FF}(\omega)$ gives also correctly the damping term entering 
\refe{FPEq}.
Thus the weak coupling calculation allows to find the form of the coefficients entering the 
Fokker-Plack equation, but it does not allow to prove the validity of the approach. 
According to the derivation presented in the main text, actually
it is not the weak-coupling condition 
that allows 
to obtain the Fokker-Plack description, but the separation of time scales. 

\bibliography{FullBib}

\end{document}